\documentclass{ws-p8-50x6-00}
\def\bn{{\bf n}}
\def\ba{{\bf a}}

\def\bP{{\bf P}}

\def\bsigma{{\bf a}}

\def\muhat{{\hat\mu}}
\def\nuhat{{\hat\nu}}
\begin{document}

\title{Exact topological density in the lattice Skyrme model\thanks{\lowercase{
\uppercase{P}resented at the \uppercase{I}nternational \uppercase{W}orkshop on \uppercase{S}trong and \uppercase{E}lectroweak \uppercase{M}atter
(\uppercase{SEWM} 2000), \uppercase{M}arseille, \uppercase{F}rance, 14--17 \uppercase{J}une 2000.
}}}

\author{Benjamin Svetitsky}
\address{School of Physics and Astronomy, Raymond and Beverly Sackler
Faculty of Exact Sciences,
Tel Aviv University, 69978 Tel Aviv, Israel\\
E-mail: bqs@julian.tau.ac.il}
\author{Alec J. Schramm}
\address{Department of Physics, Occidental College, Los Angeles,
California 90041, USA\\
E-mail: alec@oxy.edu}

\maketitle

\abstracts{
We propose using the Skyrme model on a lattice as an effective field theory
of meson--baryon interactions.
To this end we construct a local topological density that involves  the volumes 
of tetrahedra
in the target space $S^3$ and we make use of Coxeter's formula for
the Schl\"afli function to implement it.
We calculate the mean-square radius of a skyrmion in the
three-dimensional Skyrme model, and find some surprises.
}

\section{Why a lattice Skyrme model?}
The Skyrme model is a theory of a scalar field $U(x)\in SU(2)$ with the action
\begin{equation}
S=\int d{\bf x}\left[\frac{f_\pi^2}{16}\,{\rm Tr}\,|\partial_\mu U|^2
-\frac{1}{32e^2}{\rm Tr}\,
([\partial_\mu UU^{\dag},\partial_\nu UU^{\dag}]^2)\right].
\end{equation}
The action possesses an $SU(2)\times SU(2)$ chiral symmetry which is
spontaneously broken.
The Goldstone bosons are taken to represent pions, massless unless we add
a symmetry-breaking term to $S$.
The theory also contains solitons, stabilized by the 4-derivative
Skyrme term shown, that have the properties of baryons.
The model furnishes a rich phenomenology of pion--nucleon interactions
at low momenta.

The Skyrme model does not really exist as a continuum theory, since the
action is non-renormalizable.
For this reason, existing treatments of the model are semiclassical,
quantizing only the collective degrees of freedom of the soliton.
Full quantization of the theory requires a cutoff.
In considering the sector without skyrmions,
this cutoff can be removed order by order in perturbation theory at
the price of an ever-lengthening list of higher-dimension counter\-terms.
In the skyrmion sector, however, even this is difficult to accomplish.

We suggest \cite{SS}
that the limitations of the ``continuum'' model be turned to
advantage by treating the Skyrme model as an effective field theory
{\em without\/} removing the cutoff.
Now the cutoff is part of the specification of the theory.
The form of $S$ may be chosen freely, subject only to phenomenological
tests.

To write a lattice Skyrme model,\cite{Saly,DeTar} we first choose an action.
We keep the cutoff---the lattice spacing---fixed.
We apply the full panoply of lattice methods to calculate quantities
of interest, going beyond semiclassical methods and beyond perturbation
theory.
Renormalization consists of adjusting the bare couplings and the lattice
spacing to match physical parameters.
No continuum limit is necessary---or possible.

\section{Lattice topology}
Continuum solitons are classified by the winding number of the field
configuration that maps the compactified 3-space $\{R^3+\infty\}$
to the field space $SU(2)$, both of which are topologically equivalent to
the 3-sphere $S^3$.
The differential volume in the target space is given by
\begin{equation}
d\tau= \frac1{24\pi^2}
\epsilon_{ijk}\,{\rm Tr\,} (U^{-1}\partial_iU) (U^{-1}\partial_jU)
(U^{-1}\partial_kU)\,d^3r
\end{equation}
which integrates to an integer $n$.
On a lattice, we calculate this differential volume directly as follows.
\begin{enumerate}
\item We cut each lattice cube into five tetrahedra:
\vskip 10pt
\centerline{\psfig{figure=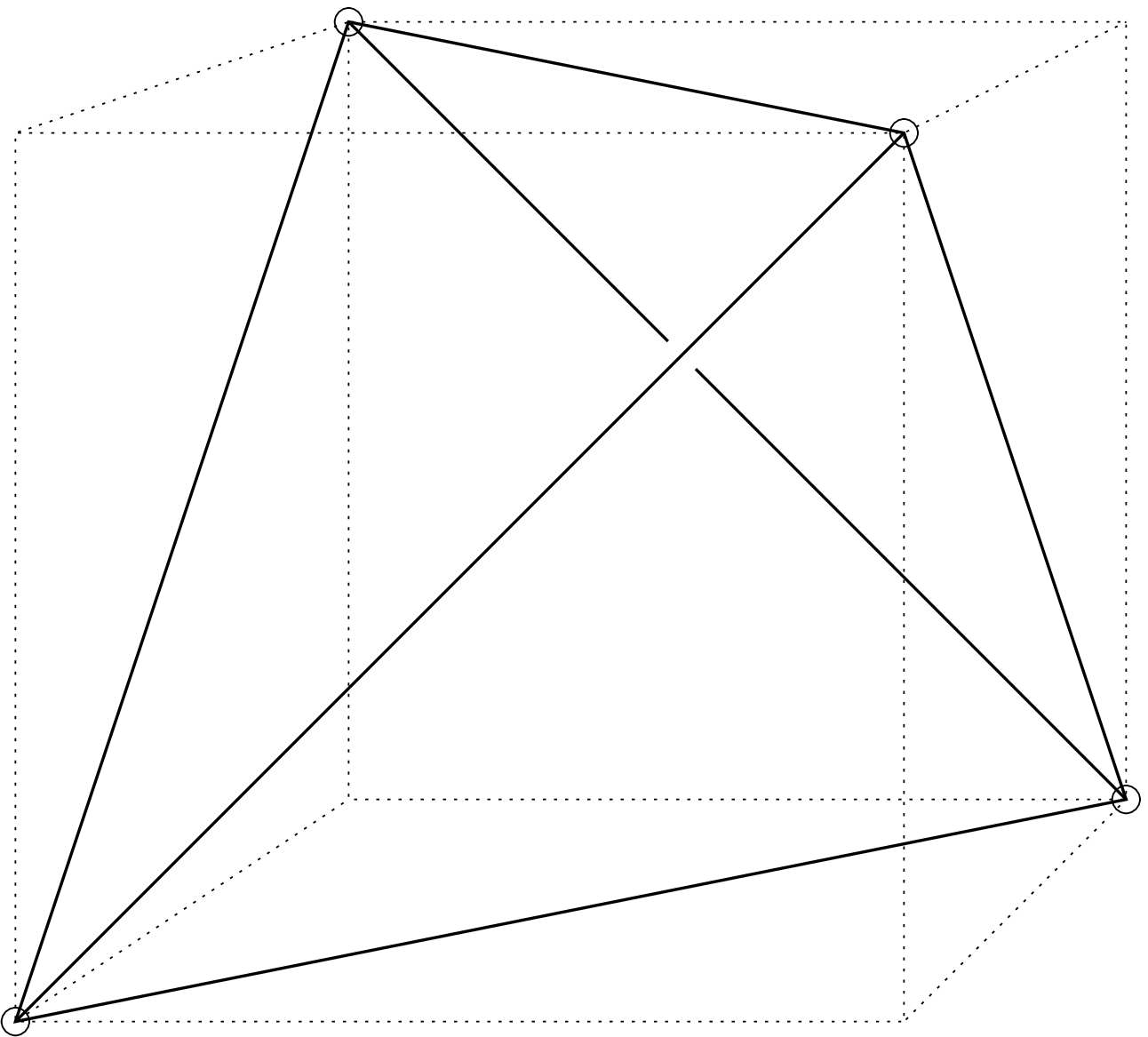,height=4cm}}
\item
Each tetrahedron maps to a tetrahedron in $S^3$.
Oddly enough, there is no simple formula for the volume of this curved
tetrahedron.\footnote{Cf.\ Girard's theorem for a
triangle in $S^2$, which says that the area is equal to the angular
excess.}
Our solution to this problem begins by dropping perpendiculars to cut
the curved tetrahedron into six {\em quadrirectangular tetrahedra\/}
(qrt's), of which we show two:

\centerline{\psfig{figure=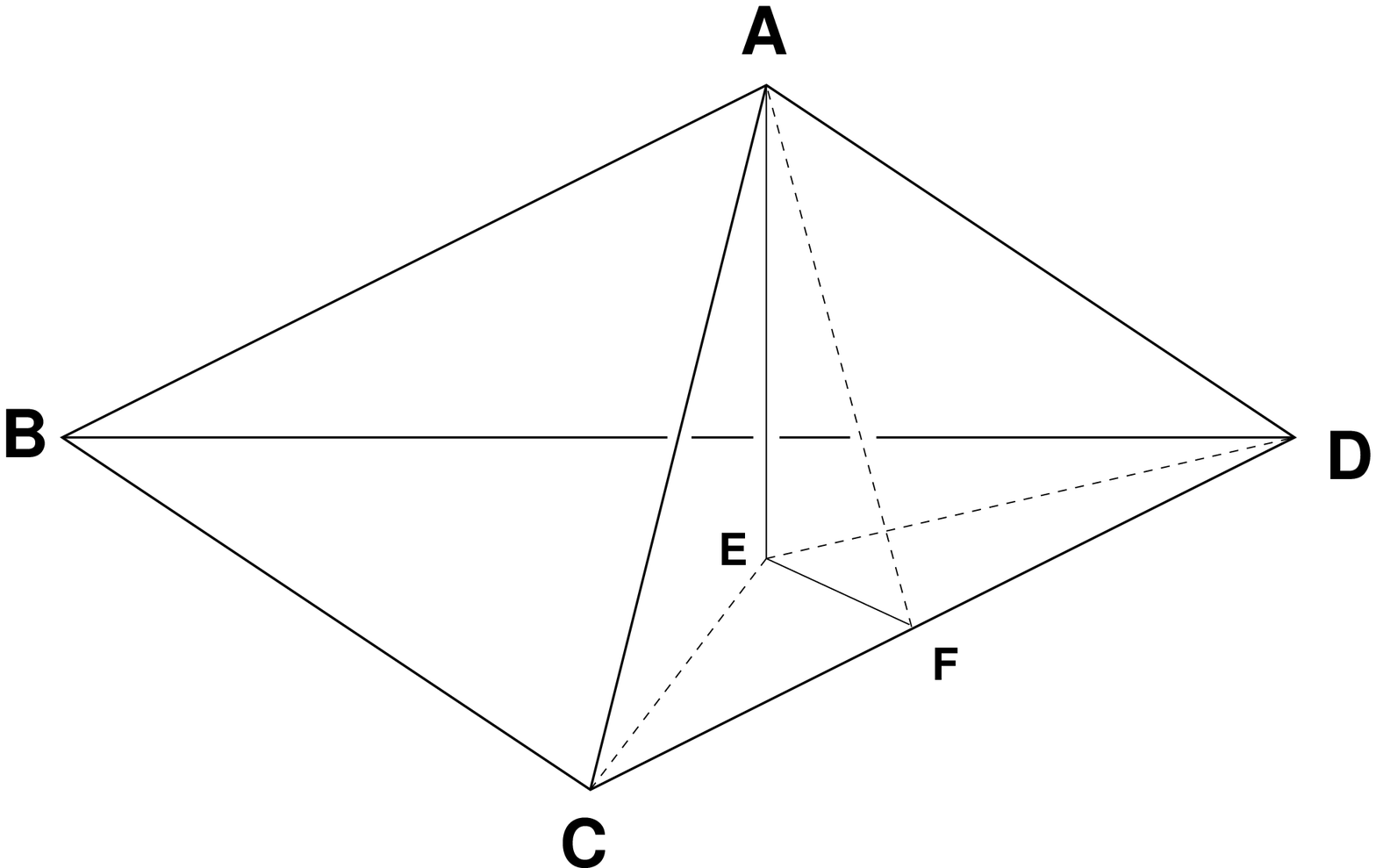,height=4cm}}
In general, a qrt can be constructed by drawing three mutually perpendicular
line segments
\def\bP{{\bf P}}
$\bP_0\bP_1\bP_2\bP_3$ and connecting all the vertices:
\vskip 10pt
\centerline{\psfig{figure=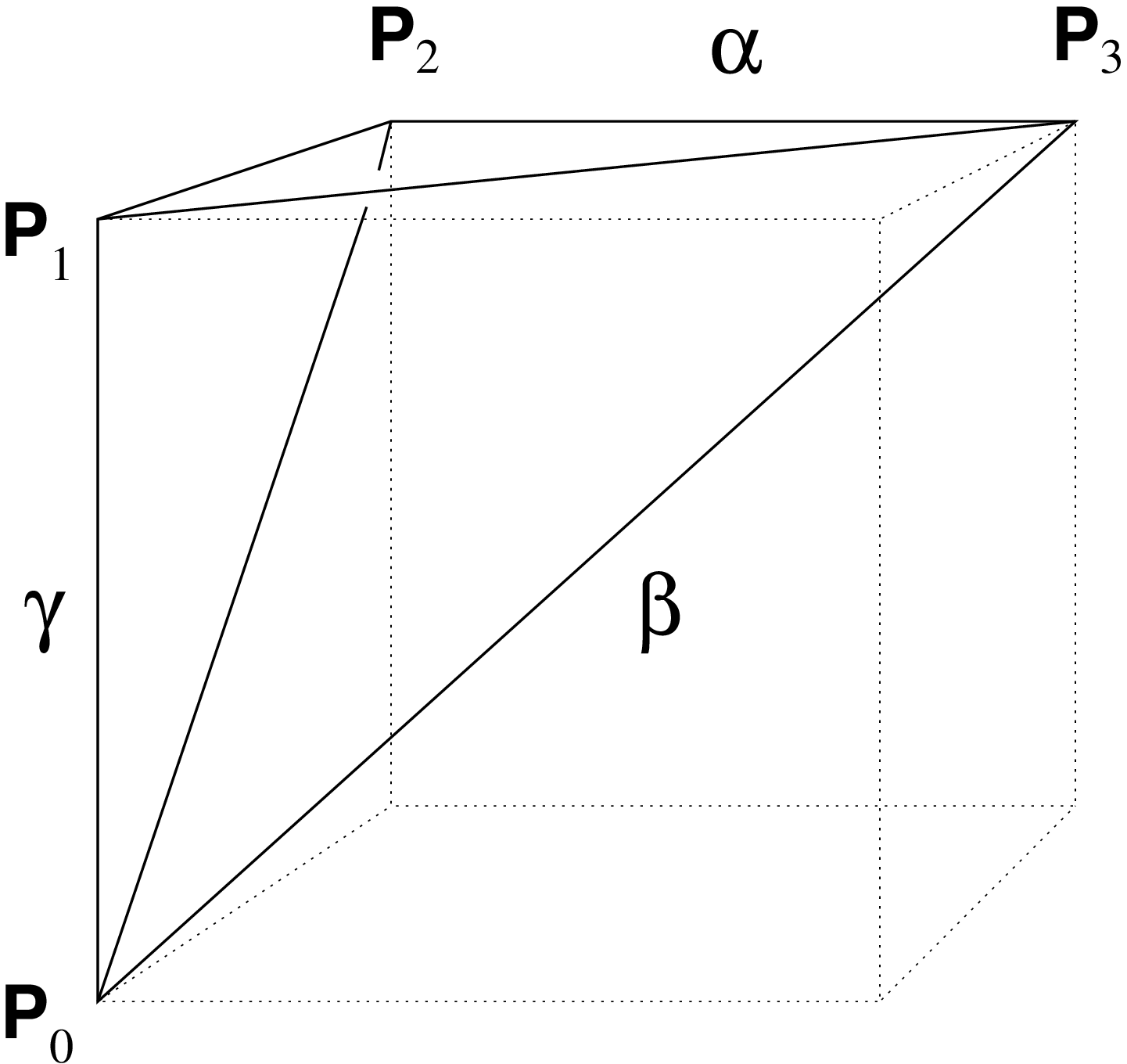,height=4.5cm}}
(This works in $S^3$ as well as well as in $R^3$.)
All the faces of a qrt are right triangles.
Three of the dihedral angles of a qrt are right angles; the other three may
be labeled $\alpha, \beta, \gamma$.
\item
Polytopes in $S^n$ were studied by Schl\"afli in the 19th century, and
in 1858 he wrote down a differential equation for the volume of
a qrt as a function of $\alpha, \beta, \gamma$.
This equation was solved by Coxeter in 1935.
The volume $V$ is given by the Schl\"afli function $S$ according to
\begin{eqnarray}
V(\alpha,\beta,\gamma)&=&\frac14S
\left(\frac\pi2-\alpha,\beta,\frac\pi2-\gamma\right)\\
S(x,y,z)&\equiv&\sum_{m=1}^\infty\left(\frac{D-\sin x\sin z}{D+\sin x\sin z}\right)^m
\frac{\cos 2mx-\cos 2my+\cos 2mz-1}{m^2}\nonumber\\
&&\qquad-x^2+y^2-z^2
\end{eqnarray}
where
$D=\sqrt{\sin^2\alpha\sin^2\gamma-\cos^2\beta}$
is the ``angular excess'' --- which vanishes for a Euclidean qrt.
In fact $V\to 0$ as $D\to0$, as expected.
\end{enumerate}

\section{The lattice action}
The topological density defined above will not be meaningful until we resolve
ambiguities associated with what is known as ``topology dropping through the
lattice.''
A field configuration $U(x)$ specifies the vertices of all the tetrahedra
in $S^3$.
Given the vertices of a tetrahedron, however, there are two ways to define
its interior: either that part of $S^3$ that includes the north pole (for
example), or that part that doesn't.
Taking one of these to be positive, say $0<V<1$ (in units of $2\pi^2$, the
volume of $S^3$), the other will be $V-1<0$.
Thus we cannot simply calculate the winding number of a given field
configuration.
We can resolve this ambiguity by {\em defining\/} the volume of a tetrahedron
to satisfy always $|V|<1/2$.

But now consider a tetrahedron whose volume is $1/2-\epsilon$.
A small change in one of the vertices can push its volume to $1/2+\delta$,
which will now be interpreted as $-1/2+\delta$, while the volumes of its
neighbors change by $\delta+\epsilon$.
Thus the winding number by our convention has jumped by 1: Topology has
fallen through the lattice.
The only way to avoid this is to force the field configurations to be sufficiently smooth that a tetrahedron volume stays well away from $\pm1/2$.
We can do this by choosing a sufficiently stiff kinetic term, such as
\cite{Ward}
\begin{equation}
S_1=(\alpha-1)\sum_{\bn\mu}\log(\bsigma_\bn\cdot\bsigma_{\bn+\muhat}-\alpha),
\label{S1}
\end{equation}
with $U=a_0+ia_i\sigma_i$.
This constrains $\ba\cdot\ba'>\alpha$.
We find that setting $\alpha=0.1$ keeps the tetrahedra well away from
ambiguity.

With the action (\ref{S1}) alone, solitons will collapse to radii on the
order of the lattice spacing before they notice that the action isn't 
quadratic in derivatives (\`a la Derrick's Theorem).
To make sizable skyrmions possible, we add a Skyrme term
\begin{eqnarray}
S_2=4\sum_\bn\sum_{\mu>\nu}&&\left\{
\left(\bsigma_{\bn+\muhat}-\bsigma_{\bn+\nuhat}\right)^2
\left(\bsigma_{\bn+\muhat+\nuhat}-\bsigma_{\bn}\right)^2\right.\nonumber\\
&&\qquad- \left.\left[
\left(\bsigma_{\bn+\muhat}-\bsigma_{\bn+\nuhat}\right)\cdot
\left(\bsigma_{\bn+\muhat+\nuhat}-\bsigma_{\bn}\right)
\right]^2\right\}.
\end{eqnarray}
The total action is $S=\beta_1S_1+\beta_2S_2$.

Will the system still tunnel between topological sectors?
In principle, yes. 
A non-local updating scheme could create a smooth skyrmion at a blow,
which would change the winding number by 1.
This won't happen with a local algorithm, such as local Metropolis
updating.

This determines how to choose an updating scheme.
Thus, if we are interested in doing thermodynamics with a chemical potential,
summing over topological sectors according to
\begin{equation}
Z(\mu)=\sum_n e^{\mu n} Z_n,
\end{equation}
we would choose a scheme that does nucleate smooth skyrmions (or even
point-like defects).
Our interest, however, is in studying the properties of a single skyrmion,
so we trap it in our computer by insisting on local updating that preserves
winding number.

\section{The Skyrme model in three dimensions, and the future}

As a first application, we have looked \cite{SS}
at the three-dimensional Skyrme model,
which may be thought of as a Ginzburg-Landau theory for the full quantum
theory at nonzero temperature.
The bare couplings in the action (at fixed cutoff) depend on physical
parameters such as $f_\pi$, the baryon mass, and the temperature, but we
have not yet tackled this matching problem.
We ran Monte Carlo simulations in the single-skyrmion sector and measured
the mean-square radius $\langle R^2\rangle$ of the topological density
defined above.
We found, to our surprise, that our lattice action $S$ admits a multitude
of metastable minima, configurations of the skyrmion that have (generally)
smaller radii than the skyrmion that actually minimizes the action.
As the ``temperature'' of the 3d model---namely, $1/\beta_1$ for fixed
$\beta_2/\beta_1$---is raised, more of these local minima come into play
and the skyrmion actually ends up {\em shrinking\/} as it is heated.
Recall that in this Ginzburg-Landau theory, this temperature reflects both
thermal and quantum fluctuations of the 4d quantum theory.
It is possible that
renormalization of the ratio $\beta_2/\beta_1$ will destroy this effect.

Directions for further study include simulation of the full 4d theory and
renormalization of its couplings (at fixed cutoff); 
study of the dependence of the size and shape of the equilibrium skyrmion on
the physical temperature and density;
and measuring the sensitivity of various results to the chosen form
of the lattice action.

\section*{Acknowledgments}

This work was supported by the Israel Science Foundation under Grant
No.~255/96-1 and by the Research Corporation.


\begin{thebibliography}{99}
\bibitem{SS}A.~J.~Schramm and B.~Svetitsky,
%``Topology and metastability in the lattice Skyrme model,''
hep-lat/0008003.
%%CITATION = HEP-LAT 0008003;%%
\bibitem{Saly}R.~Saly,
%``Lattice Skyrme Model,''
Phys.\ Rev.\ D {\bf 31}, 2652 (1985).
%%CITATION = PHRVA,D31,2652;%%
\bibitem{DeTar}C.~DeTar,
%``The Role Of Baryons In Chiral Symmetry Restoration At High Temperature,''
Phys.\ Rev.\ D {\bf 42}, 224 (1990).
%%CITATION = PHRVA,D42,224;%%
\bibitem{Ward}R.~S.~Ward,
%``Stable topological skyrmions on the 2-D lattice,''
Lett.\ Math.\ Phys.\  {\bf 35}, 385 (1995)
[hep-th/9502048].
%%CITATION = HEP-TH 9502048;%%
\end{thebibliography}
\end{document}